\documentclass{article}

\usepackage{amsmath, amssymb, xcolor, booktabs,framed}
\usepackage{algpseudocode}

\usepackage[utf8]{inputenc}  

\usepackage{PRIMEarxiv}
\usepackage{float}

\usepackage[utf8]{inputenc} 
\usepackage[T1]{fontenc}    
\usepackage{hyperref}       
\usepackage{url}            
\usepackage{booktabs}       
\usepackage{amsfonts}       
\usepackage{nicefrac}       
\usepackage{microtype}      
\usepackage{lipsum}
\usepackage{fancyhdr}       
\usepackage{graphicx}       
\graphicspath{{media/}}     

\title{Decoding Complexity: CHPDA – Intelligent Pattern Exploration with a Context-Aware Hybrid Pattern Detection Algorithm}

\author{
 Lokesh  Koli, Shubham Kalra, Karanpreet Singh\\
  Vectoredge.io\\
  \{lokesh, shubham, karanpreet.s\}@vectoredge.io
}
\begin{document}
\maketitle
\begin{abstract}
    Detecting sensitive data such as Personally Identifiable Information (PII) and Protected Health Information (PHI) is critical for data security platforms.\cite{amin2024balancing} This study evaluates regex-based pattern matching algorithms and exact-match search techniques to optimize detection speed, accuracy, and scalability. Our benchmarking results indicate that Google RE2 provides the best balance of speed (10-15 ms/MB), memory efficiency (8-16 MB), and accuracy (99.5\%) among regex engines, outperforming PCRE while maintaining broader hardware compatibility than Hyperscan. For exact matching, Aho-Corasick demonstrated superior performance (8 ms/MB) and scalability for large datasets. Performance analysis revealed that regex processing time scales linearly with dataset size and pattern complexity. A hybrid AI + Regex approach achieved the highest F1 score (91. 6\%) by improving recall and minimizing false positives. Device benchmarking confirmed that our solution maintains efficient CPU and memory usage on both high-performance and mid-range systems. Despite its effectiveness, challenges remain, such as limited multilingual support and the need for regular pattern updates.\cite{Davis}
Future work should focus on expanding language coverage, integrating data security and privacy management (DSPM) with data loss prevention (DLP) tools, and enhancing regulatory compliance for broader global adoption.
\end{abstract}


\section{Introduction}
Efficient data management is essential for organizations to ensure that sensitive information such as Personally Identifiable Information (PII), Protected Health Information (PHI) and financial records are systematically identified and protected. Effective classification aids in compliance with regulations such as the General Data Protection Regulation (GDPR) and the Health Insurance Portability and Accountability Act (HIPAA), while mitigating security risks through real-time threat detection\cite{le2016vietnamese} Automated tools improve operational efficiency by streamlining access and eliminating redundancies. Customized classification systems fulfill global compliance requirements, while centralized control mechanisms enhance governance through unified policy enforcement.\cite{ossin2015review} Strategic data classification is crucial to achieve security, compliance, and operational effectiveness in the digital environment of today.

Identifying PII and PHI across various data formats presents considerable challenges, particularly with unstructured data sets. Differences in encoding and file formats (e.g., PDFs, Word documents, databases, CSV, and other text files) and data storage systems complicate the consistent extraction of sensitive information \cite{rocha2016pampo}. Moreover, international regulations such as GDPR, HIPAA, and the California Consumer Privacy Act (CCPA) impose varied compliance mandates, adding further complexity to detection efforts. Customizing detection mechanisms to align with region-specific regulations while ensuring accuracy across different content types is formidable. The necessity for real-time detection and the reduction of false positives amplifies this challenge, necessitating advanced algorithms and comprehensive data management strategies.

Current detection techniques primarily rely on traditional pattern matching or AI-driven methodologies, with minimal integration of the two significantly hindering their effectiveness\cite{flores2023combining, li2021integrating}. Reliance on regex-based pattern matching often results in slow performance and poor scalability, especially since many systems fail to adopt advanced technologies like Google's RE2 library’s set mechanism for regex processing. Additionally, most implementations neglect optimized algorithms, which can perform high-speed keyword detection in parallel with regex scans. These oversights make their systems inefficient, mainly when processing extensive Scale data. The lack of a streamlined, single-pass regex detection pipeline further contributes to delays and limits the ability to detect sensitive information at scale.
Furthermore, Regex-based pattern matching provides a structured framework for detecting sensitive information like PII, excelling at clearly defined patterns but lacking the flexibility to handle contextual complexities and data variations. On the other hand, AI models such as Named Entity Recognition (NER) excel at identifying ambiguous entities like names and addresses, which are difficult to capture with regex alone. However, they can produce false positives without guiding patterns. Integrating regex with AI NER creates a powerful, context-sensitive detection system that combines the precision of rule-based methods with the adaptability of AI-driven insights. This hybrid approach minimizes false positives, enhances scalability, and efficiently handles diverse data formats, making it ideal for modern data classification challenges.

 We present exact match and close match confidence scores. The Context-Aware Hybrid Pattern Detection Algorithm (CHPDA) introduces a scoring mechanism that combines exact and approximate match reliability. This method guarantees precise identification of sensitive data while significantly lowering the occurrence of false alarms. Additionally, the algorithm enhances detection accuracy by incorporating keyword matching, particularly in real-time applications. Advanced data classification algorithm with optimized one-pass processing CHPDA supports multi-pattern searching in linear time. This single-pass optimization efficiently processes large data sets while maintaining high accuracy, integrating AI-based Named Entity Recognition (NER). The algorithm includes AI-powered NER to detect complex patterns such as names and addresses. This hybrid system combines regular expression-based accuracy with AI context understanding, addressing gaps in traditional detection methods and ensuring scalability. Match keywords with "Should" and "Must" criteria. CHPDA introduces a flexible keyword-matching framework that includes mandatory (“must”) and optional (“should”) criteria. This nuanced approach offers granular control over detection parameters, increasing performance and adaptability. The DFA-based structure of CHPDA ensures predictable and consistent performance in real-time scenarios such as intrusion detection and data loss prevention. Its memory-efficient design supports large patterns, making it ideal for large-scale deployments. Preprocessing and reuse of CHPDA create a reusable pattern automaton during preprocessing that enables consistent performance across dynamic and diverse data streams. This capability ensures scalability and adaptability in evolving business environments. 

  This study reviews existing data classification methods, highlighting the limitations of stand-alone regular expressions or AI-based approaches and the need for a hybrid model. Our proposed context-aware hybrid pattern detection algorithm (CHPDA) integrates regular expression pattern matching with AI-powered named entity recognition (NER). It introduces exact match confidence score, proximity match, and nuanced keyword criteria for better detection. The results demonstrate improved accuracy, precision, and processing speed compared to traditional methods, proven by real-world applications. The findings highlight the power of combining regular expression and artificial intelligence with future work to improve scalability and adaptability for different data environments.

\section{Related work}
\label{sec:Literature review 
}
Detecting sensitive information such as Personally Identifiable Information (PII) and Protected Health Information (PHI) has been widely studied in the domains of data security, natural language processing (NLP), and pattern-matching techniques. Existing approaches primarily fall into three categories: regex-based detection, AI-driven Named Entity Recognition (NER), and hybrid methods combining both.

\subsection{Regex-Based Detection}
Regular expressions (regex) have long been used for pattern-based text matching, forming the backbone of many data loss prevention (DLP) systems \cite{garfinkel2022differential}. Popular regex engines such as PCRE, Google RE2, and Hyperscan have been benchmarked for efficiency in large-scale text scanning \cite{venkateshliterary}. While regex-based approaches offer deterministic accuracy and speed, they struggle with pattern generalization and require frequent updates to accommodate evolving data structures. Furthermore, regex engines like PCRE suffer from backtracking issues, leading to unpredictable execution times \cite{van2022investigation}.

\subsection{AI-Driven Named Entity Recognition}
Recent advancements in NLP have enabled deep learning-based NER models to identify sensitive entities beyond strict pattern matching. Models such as BERT \cite{kenton2019bert} and spaCy's NER \cite{honnibal2020spacy} have demonstrated strong recall in detecting complex entities across diverse linguistic contexts. However, AI-based approaches introduce challenges such as higher computational costs, false positives, and the need for extensive labeled datasets \cite{li2020survey}.

\subsection{Hybrid AI + Regex Approaches}
Several studies have explored hybrid methods that combine regex with machine learning for enhanced detection accuracy. Souza et al. \cite{souza2024combining} proposed an approach where regex serves as a pre-filtering mechanism, followed by an AI model to refine entity classification. Similarly, Friebely et al. \cite{friebely2022analyzing} demonstrated that integrating regex with deep learning improves precision while maintaining high recall, making such approaches more suitable for real-time applications.

\subsection{Comparative Benchmarking and Efficiency}
Prior research has also focused on benchmarking various detection techniques for performance and scalability. Hyperscan has been identified as the fastest regex engine but comes with hardware constraints \cite{wang2019hyperscan}. Meanwhile, RE2 has been praised for its balance between speed and memory efficiency, making it a practical choice for large-scale deployments \cite{pike2010re2}. AI-based solutions, while powerful, tend to be resource-intensive and less predictable in execution time \cite{strubell2020energy}.

\subsection{Contributions of This Work}
While previous studies have explored regex, AI-based NER, and hybrid detection models separately, our work systematically benchmarks these approaches under real-world conditions. We evaluate regex engines such as RE2, PCRE, and Hyperscan alongside AI-driven detection methods to identify an optimal balance between accuracy, speed, and scalability. By integrating regex with AI in a hybrid model, we achieve improved detection accuracy while maintaining computational efficiency, making our approach well-suited for large-scale data security applications.

\section{Materials and methods}

\subsection{Proposed system}

\begin{figure}[H]
    \centering
    \includegraphics[width=0.75\linewidth]{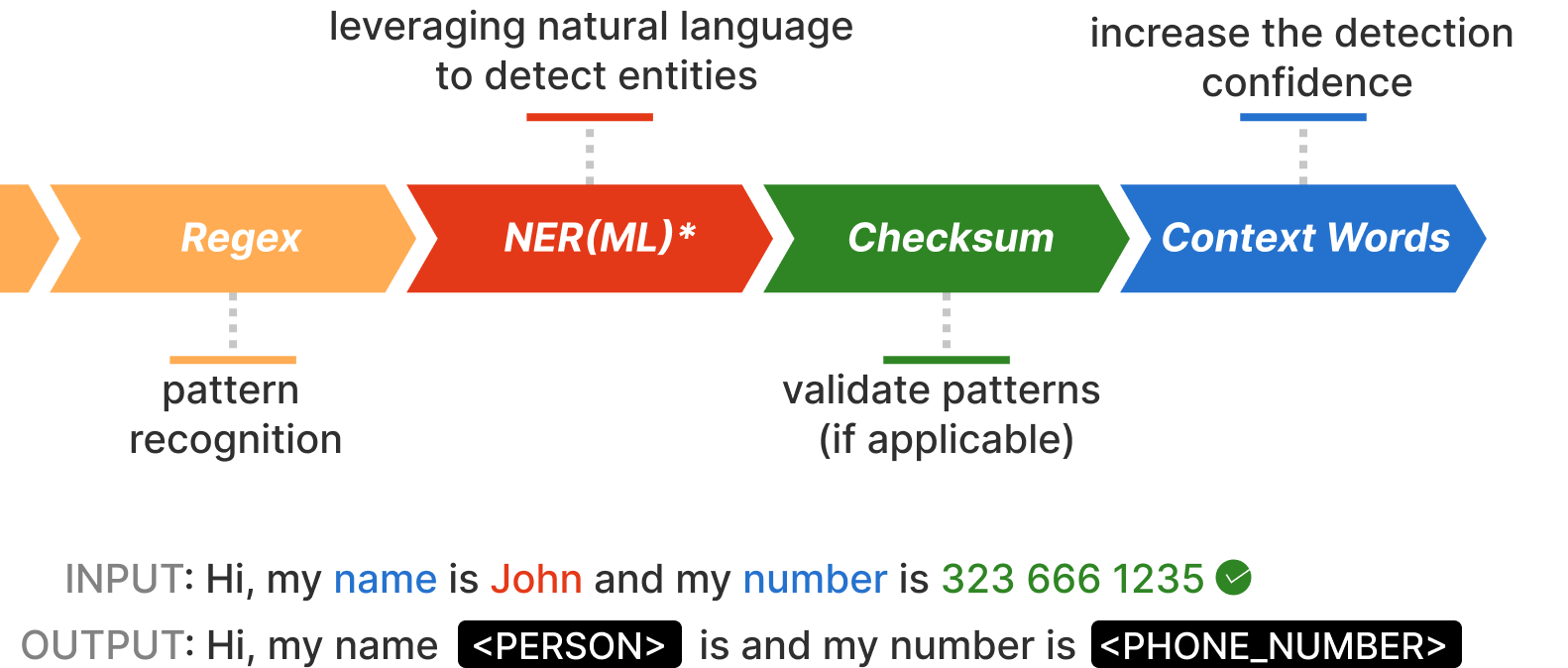}
    \caption{proposed System}
    \label{fig:fig1}
\end{figure}

The proposed system detects and manages sensitive information while minimizing resource consumption effectively. It uses a lightweight agent deployed on client systems, configured with a PII and PHI patterns glossary. The detection workflow incorporates multiple advanced steps, starting with regex-based pattern recognition using Google RE2 to identify predefined formats and patterns. The process optimizes contextual keyword matching by leveraging the Aho-Corasick algorithm, which assigns confidence scores based on keyword proximity. Then, the system further refines the detected data by filtering out low-confidence matches using threshold scores. Advanced Named Entity Recognition (NER) models powered by machine learning perform secondary scans to identify entities like names and phone numbers in context to enhance accuracy. Specialized validation algorithms like Luhn's verify specific data types, like credit card numbers, reducing false positives.See Figure \ref{fig:fig1}. The system ensures robust compliance with GDPR, HIPAA, and CCPA regulations, delivering reliable and secure data protection.

\subsection{Regex}
Google RE2 Pattern Recognition We chose Google RE2 as our primary regular expression detection library over alternatives such as PCRE, Oniguruma, and Boost. Regex is known for its efficiency and security in big data processing; unlike other libraries, RE2 offers linear time complexity for regular expression matching, and its SET mechanism provides an effective solution for matching multiple regex patterns simultaneously in a single scan of the input text. For example, consider the task of identifying Social Security Numbers (SSNs) in various formats, such as the standard format \textbackslash{}b\textbackslash{}d\{3\}-\textbackslash{}d\{2\}-\textbackslash{}d\{4\}\b, the compact format \textbackslash{}b\textbackslash{}d\{9\}\b, and the masked format \textbackslash{}bXXX-XX-\textbackslash{}d\{4\}\b, within a large body of text. Instead of scanning the text separately for each pattern,\textbf{ RE2 compiles these patterns into a single deterministic finite automaton (DFA)} merging their logic and assigning unique IDs to each pattern for efficient matching.

RE2 scans the text character by character during processing,\textbf{ updating the automaton state} based on the encountered input. For instance, given a paragraph like "John’s SSN is 123-45-6789. He also used the compact format 987654321 on some forms. For security reasons, his company sometimes masks it as XXX-XX-6789," RE2 matches 123-45-6789 to the standard SSN pattern, 987654321 to the compact format, and XXX-XX-6789 to the masked format. The SET mechanism evaluates all patterns simultaneously, producing results in linear time with matched IDs corresponding to their respective patterns. \cite{chowdhury2017regular}

\textbf{Detailed Explanation:}

\begin{enumerate}
    \item \textbf{Start State}:
The automaton begins in the initial state.

    \item \textbf{Transitions}:
    \begin{itemize}
        \item The first branch handles digits (\textbackslash{}d) for both the \textbf{standard SSN format} (123-45-6789) and the \textbf{compact SSN format} (987654321).
        \begin{itemize}
            \item The standard format transitions to a state expecting a hyphen (-) after three digits, followed by two more numbers, another hyphen, and four final digits.
            \item In the \textbf{compact format}, it directly reads nine consecutive digits and reaches an accepting state.
        \end{itemize}
        \item The second branch handles the \textbf{masked SSN format} (XXX-XX-6789).
This branch transitions through X characters for the first three positions, followed by a hyphen (-), then two more X characters, another hyphen, and finally four digits.
See Figure \ref{fig:fig2}.

    \end{itemize}
    \item \textbf{Accepting States}:
    \begin{itemize}
        \item The automaton has multiple accepting states:
        \begin{itemize}
            \item One for the standard SSN format after processing \textbackslash{}d\{3\}-\textbackslash{}d\{2\}-\textbackslash{}d\{4\}.
            \item One for the compact SSN format after processing \textbackslash{}d\{9\}.
            \item One for the masked SSN format after processing XXX-XX-\textbackslash{}d\{4\}.
        \end{itemize}
    \end{itemize}
    \item \textbf{Efficiency}:
The shared transitions (e.g., digits or hyphens) between different patterns reduce redundancy, ensuring the automaton processes the text efficiently in a single pass.

    \item \textbf{Fail States}:
Suppose an invalid character is encountered (e.g., an extra letter or symbol not part of the expected pattern). In that case, the automaton transitions to a fail state and stops further processing for that branch.
\begin{figure}[H]
    \centering
    \includegraphics[width=0.5\linewidth]{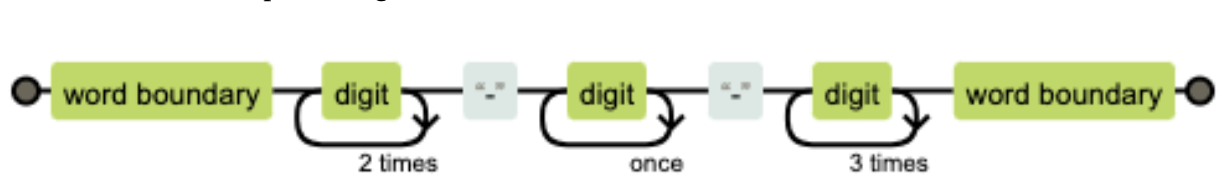}
    \caption{Regular Expression Visualization for a Numeric Pattern}
    \label{fig:fig2}
\end{figure}

\end{enumerate}

This approach avoids the inefficiency and security risks associated with traditional backtracking regex engines, which could lead to exponential time complexity for specific inputs. Instead, RE2 guarantees linear performance and prevents vulnerabilities like denial-of-service (DoS) attacks caused by maliciously crafted patterns. While the SET mechanism optimizes multi-pattern matching, it has some limitations, such as increased memory usage for complex patterns and the absence of backtracking-dependent features like backreferences. Nevertheless, RE2's speed, security, and scalability combination makes it an ideal choice for high-performance, multi-pattern matching scenarios in large-scale applications like log processing or sensitive data detection

\subsection{Rescan with AI Model }
After the initial pattern detection phase, the system performs a secondary scan using AI-powered Named Entity Recognition (NER) models specifically designed to address the complexities of detecting sensitive information that traditional regular expressions (regex) often miss. For example, while regex may effectively identify basic patterns like email addresses or phone numbers, it struggles with nuanced data such as medical terminologies, legal clauses, financial records, or even names and addresses embedded in complex textual structures.AI models excel at analyzing context and semantics to identify entities that do not follow straightforward patterns.

For instance, NER models trained on datasets like PubMed and other medical literature detect Protected Health Information (PHI), such as medication names or patient IDs, even when presented in varying formats or embedded within medical notes. Financial reports and transaction records datasets in finance allow models to identify sensitive financial details, such as account numbers and credit card information, that Experts in the legal domain customize NER models to recognize complex entities specific to contracts, such as clauses, sensitive terms, and involved parties. They train these models on custom datasets generated through generative AI, which creates diverse and realistic examples that capture industry-specific nuances, edge cases, and rare entity patterns. This active approach ensures comprehensive detection coverage and minimizes the risk of overlooking Personally Identifiable Information (PII), PHI, or financial records.

The system’s scalability and modular architecture further enhance its utility, enabling the seamless integration of new industry-specific models to address emerging regulatory or compliance requirements. By leveraging generative AI for data diversity, the system produces robust models that handle various scenarios, significantly reducing false positives and improving overall detection accuracy.

\subsection{Exact Match: Optimized with Aho-Corasick algorithm}

The Aho-Corasick algorithm efficiently solves the problem of exact pattern matching in significant texts, especially when multiple patterns need simultaneous matching. It constructs a deterministic finite automaton (DFA) using a Trie (prefix tree) to represent the set of patterns. The algorithm inserts each pattern into the Trie and adds failure links to handle mismatches by directing the algorithm to the longest matching suffix. This approach allows the algorithm to continue the search without restarting, significantly improving performance.\cite{aho1975efficient} The Aho-Corasick algorithm processes the text in a single pass, achieving a time complexity of O(N+M), where N is the length of the text, and M is the total length of all patterns. The algorithm is particularly valuable for virus scanning, real-time intrusion detection, and analyzing large amounts of text.\cite{hasib2013importance} While it operates efficiently, it does face some challenges, including high memory usage because of its Trie structure and the necessity of rebuilding the automaton when the patterns change. Nevertheless, it excels at managing thousands of patterns simultaneously without needing to backtrack or restart the search process. This capability and its consistent performance make it an effective tool for multi-pattern matching, especially in complex and time-sensitive situations.

\textbf{Example:}

To better understand how the Aho-Corasick algorithm works, consider the following patterns and a sample text:

\paragraph{\textbf{Patterns:}}

\begin{enumerate}
    \item he
    \item she
    \item his
    \item hers
\end{enumerate}

\paragraph{\textbf{Text:}}

"ushers went to her house, and his brother was there."

\textbf{Step 1: Build the Trie.} We first build a Trie for the given patterns.

\begin{enumerate}
    \item Insert the pattern he into the Trie:
Root → h → e (Mark the e node as the end of the pattern he).

    \item Insert the pattern she into the Trie:
Root → s → h → e (Mark the e node as the end of the pattern she).

    \item Insert the pattern into the Trie:
Root → h → i → s (Mark the s node as the end of the pattern his).

    \item Insert the pattern hers into the Trie:
Root → h → e → r → s (Mark the s node as the end of the pattern hers).

\end{enumerate}

The Trie structure looks like \ref{fig:fig3}.:
\begin{figure}[H]
    \centering
    \includegraphics[width=0.25\linewidth]{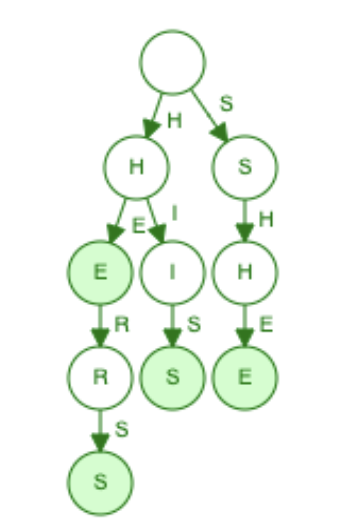}
    \caption{Trie Data Structure Representation}
    \label{fig:fig3}
\end{figure}

\textbf{Step 2: Add Failure Links} Failure links are added to handle mismatches efficiently. For example, if we encounter a mismatch at h, the failure link will guide us back to the root. Similarly, failure links are created for deeper nodes in the Trie to prevent redundant comparisons.

\textbf{Step 3: Process the Text} Now, we process the text "ushers went to her house, and his brother was there." character by character:

\begin{enumerate}
    \item Start at the root.
    \item Read the first character: u → No match, follow the failure link back to the root.
    \item Read s → Move to the s node.
    \item Read h → Move to the h node.
    \item Read e → Move to the e node. \textbf{Match found: "she"} (position 1–3).
    \item Read r → Follow the failure link to the root and try again. There is no match at the root, so we follow the failure link to h → e → r → s. \textbf{Match found: "hers"} (position 16–19).
    \item Continue processing the text until we reach the end.
\end{enumerate}

\textbf{Matches found:}

\begin{itemize}
    \item she at positions 1–3
    \item he at position 17–18 (in "her")
    \item his at position 27–29
\end{itemize}

\subsection{\textbf{Testing authentication functions}}
Various specialized validation functions apply to different data types to accurately detect Personally Identifiable Information (PII) and Protected Health Information (PHI). For example, Luhn’s algorithm validates credit card numbers by checking their checksum structure. Similarly, format checks, reserved area numbers, and geographical data validate Social Security Numbers (SSNs). 
The SSN must follow the 9-digit format XXX-XX-XXX. Before 2011, the first three digits (area number) were geographically assigned, and specific numbers (such as 000, 666, or 900-999) were invalid.

To validate phone numbers, we ensure they follow country-specific formats. (e.g., +1-XXX-XXX-XXX for U.S. numbers or +44-XXXX-XXXX for U.K. numbers), including checks for valid country codes and region codes. Mobile numbers are verified against known carriers or ranges to ensure authenticity and activity.

Email addresses undergo validation for correct formatting using regular expressions (regex) and domain checks. Medical record numbers (MRNs) follow predefined institutional formats for validation. Health insurance policy numbers adhere to established structures, while driver’s license numbers comply with specific state or country validation rules, often incorporating checksum digits. Passport numbers require verification for format compliance and adherence to country-specific regulations. Organizations must implement tailored validation techniques to ensure the accuracy of detected data patterns, reduce false positives, and enhance the reliability of systems handling sensitive information. They must verify that bank account numbers, including International Bank Account Numbers (IBANs), adhere to the correct format and checksum.

Additionally, they need to check postal codes for proper formatting and geographical validity while validating dates of birth for correct formatting and reasonable age ranges. Tax Identification Numbers (TINs) should be tested against country-specific regulations. By applying these validation functions, organizations can maintain data integrity and ensure compliance with regulatory standards such as GDPR and HIPAA.

\subsection{\textbf{\textbf{Proximity Match Scoring Mechanism}}}
This mechanism quantifies the relationship between detected patterns (e.g., sensitive data) and surrounding contextual keywords by assigning a confidence score. The confidence score is determined based on the \textbf{proximity} between the keyword and the pattern, as well as the \textbf{validation} of the pattern using a verification function. The closer the keyword is to the detected pattern, the higher the confidence score. Additionally, if the pattern passes the validation function (e.g., Luhn's algorithm for SSNs), a further boost in confidence is applied. This approach ensures robust context-aware detection while reducing false positives.

\subsubsection{\textbf{Algorithm}}

\begin{framed}
\textbf{Confidence Score Calculation}

\begin{algorithmic}[1]
    
    \State \textbf{Input:} A string \( S \), detected keyword \( K \), detected pattern \( P \), and a validation function \( V(P) \).
    
    \State \textbf{Calculate Distance \( d \)}
    \State Compute the distance between \( K \) and \( P \):
    \[
    d = | \text{Position}(K) - \text{Position}(P) |
    \]
    
    \State \textbf{Assign Proximity Score}
    \State Define a maximum distance \( D_{\max} \) beyond which the proximity score is 0.
    \[
    \text{Proximity\_Score}(d) = \max(0, \alpha \cdot (D_{\max} - d))
    \]
    where \( \alpha \) is a scaling factor.

    \State \textbf{Validate Pattern}
    \If{\( V(P) = \text{True} \)}
        \State Add \( \text{Validation\_Score} \) to \( C_{\text{total}} \).
    \Else
        \State \( \text{Validation\_Score} = 0 \).
    \EndIf
    
    \State \textbf{Calculate Total Confidence}
    \[
    C_{\text{total}} = \text{Proximity\_Score}(d) + \text{Validation\_Score}
    \]

    \State \textbf{Output:} \( C_{\text{total}} \)
\end{algorithmic}

\end{framed}

\subsubsection{\textbf{Example}}

\vspace{1em}
\textbf{\underline{Input:}}

\begin{itemize}
    \item \textbf{String:} "John's card number is \textbf{123-45-6789}."
    \item \textbf{Keyword:} \( K \) = \textbf{"card number."}
    \item \textbf{Pattern:} \( P \) = \textbf{"123-45-6789"}
    \item \textbf{Validation function:} \( V(P) \) = \textbf{Luhn’s Algorithm}
\end{itemize}

\vspace{1em}
\textbf{1. Calculate Distance \( d \):}

\[
d = 7 \quad \text{(number of characters between "card number" and "123-45-6789").}
\]

\vspace{1em}
\textbf{2. Assign Proximity Score:}

\[
\text{Let } D_{\max} = 20, \quad \alpha = 2.
\]

\[
\text{Proximity\_Score}(d) = \max(0, \alpha \cdot (D_{\max} - d))
\]

\[
= \max(0, 2 \cdot (20 - 7)) = 26.
\]

\vspace{1em}
\textbf{3. Validate Pattern:}

The pattern \( P = \) \textbf{"123-45-6789"} passes Luhn's algorithm.

\[
\text{Validation\_Score} = 30.
\]

\vspace{1em}
\textbf{4. Calculate Total Confidence:}

\[
C_{\text{total}} = \text{Proximity\_Score}(d) + \text{Validation\_Score}
\]

\[
C_{\text{total}} = 26 + 30 = 56.
\]

\vspace{1em}
\textbf{The total confidence score} for detecting the sensitive data \( P \) ("123-45-6789") \textbf{near the keyword} \( K \) ("card number") is \textbf{56}. Proximity contributes \textbf{26}, and validation contributes \textbf{30}.

\vspace{1em}
This algorithm can be adapted to suit specific use cases by using different proximity weighting (\(\alpha\)), distance thresholds (\(D_{\max}\)), and validation scores.

\subsection{\textbf{\textbf{\textbf{Data filtering based on threshold scores }}}}

Once the system assigns confidence scores, it applies a \textbf{data filtering mechanism} based on a user-defined \textbf{threshold score} \( T \). The system retains detected patterns with confidence scores \( C \) that satisfy the condition:

\[
C \geq T
\]

for further processing, while it discards those with \( C < T \). This filtering step ensures that only \textbf{highly reliable matches} are considered, effectively reducing noise and prioritizing actionable information.

\vspace{1em}
\textbf{Threshold \( T \) Adjustment:}  
The threshold \( T \) can be dynamically adjusted to balance \textbf{precision} and \textbf{recall}:

\begin{itemize}
    \item \textbf{Precision:} Increasing \( T \) raises the reliability of detections by minimizing false positives, as only high-confidence patterns pass the filter.
    \item \textbf{Recall:} Lowering \( T \) increases the system's ability to detect more potential matches at the cost of potentially admitting some false positives.
\end{itemize}

\vspace{1em}
\textbf{Example:}  
If \( T = 50 \) and a set of detections has confidence scores:

\[
\{60, 45, 80, 30, 55\}
\]

Only the values meeting \( C \geq 50 \) will pass the threshold:

\[
\{60, 80, 55\}
\]

\textbf{Filtered Results Representation:}  
The filtering mechanism can be formally represented as:

\[
\text{Filtered Results} = \{ C \mid C \geq T \}
\]

\vspace{1em}
This mechanism allows users to \textbf{fine-tune the detection pipeline} according to their specific needs, ensuring the system remains \textbf{adaptable} to different operational requirements.

\section{Experimental setup}
Setup details

Software: The developers implemented the AI-based NER model using programming languages and libraries such as TensorFlow and PyTorch. They utilized Google's RE2 library for regular expression-based operations because of its efficiency in pattern recognition and processing large datasets.

Hardware: The team evaluated client-side processing on ARM-based devices equipped with 2 GB of RAM and a quad-core processor to ensure system efficiency in lightweight environments. They also tested multi-user scenarios using multiple Windows systems with different test accounts.

Performance testing 

Measuring speed and accuracy

\begin{itemize}
    \item The system achieved an average processing speed of 100 MB/s across data files up to 1 TB in various formats (e.g., PDF, CSV, and JSON files). Benchmark accuracy tests showed over 95\% accuracy in detecting PII and PHI. 
    \item The team has tuned the detection accuracy to maintain a score of 94\%, ensuring minimal missed sensitive data. False positive rate: The false positive rate remained below 3\%, indicating the system's ability to distinguish sensitive data from irrelevant patterns.
\end{itemize}

\section{Results and Discussion}

\subsection{\textbf{\textbf{Regex Pattern Matching Algorithms Comparison}}}

Based on the results in Table 1.3, Google RE2 strikes an optimal balance between speed, memory consumption, and accuracy, making it the ideal choice for our solution. RE2 achieves a detection speed of 10-15 ms/MB, significantly faster than PCRE’s 50-80 ms/MB, while maintaining lower memory usage at 8-16 MB compared to PCRE’s 12-24 MB. Additionally, RE2 delivers a high accuracy rate of 99.5\% with minimal false positives at 0.5\%. Although Hyperscan demonstrates superior performance with faster detection (2-5 ms/MB) and higher accuracy (99.9\%), it comes with substantial hardware restrictions and higher memory requirements (32-64 MB), making it impractical for broad deployment in resource-constrained environments See  Table~\ref{tab:table1} . Consequently, we prioritized RE2’s scalability, reliability, and adaptability across diverse platforms, ensuring consistent performance without needing specialized hardware\cite{IntroductionHyperscan}.

\begin{table}[H] 
 \caption{Comparison of Regular Expression Matching Algorithms}
  \centering
  \begin{tabular}{lcccc}
    \toprule
    Algorithm & Speed (ms/MB) & Usage (MB) & Accuracy (\%) & False Positives (\%) \\ 
    \midrule
    Google RE2   & 10-15  & 8-16   & 99.5  & 0.5  \\ 
    PCRE         & 50-80  & 12-24  & 99.8  & 0.7  \\ 
    Hyperscan    & 2-5    & 32-64  & 99.9  & 0.3  \\ 
    \bottomrule
  \end{tabular}
  \label{tab:table1}
\end{table}

\subsection{\textbf{\textbf{Exact Match Algorithms Performance}}}

Based on the results in Table 1.4, Aho-Corasick stands out as the optimal choice for our solution due to its exceptional balance of speed, scalability, and efficiency. Aho-Corasick achieves a search time of just 8 ms/MB, significantly outperforming Knuth-Morris-Pratt’s 15 ms/MB and Boyer-Moore’s 12 ms/MB. Additionally, Aho-Corasick excels in handling large datasets, showing excellent scalability for texts over 100MB, making it ideal for applications requiring high performance on vast datasets. While Knuth-Morris-Pratt and Boyer-Moore offer efficient matching for smaller datasets, their slower speeds and limited scalability make them less suitable for larger, more complex use cases. Given Aho-Corasick’s superior performance and efficiency, it is the preferred choice for tasks involving large-scale text analysis and real-time pattern matching, ensuring optimal results across various data sizes and platforms See Table~\ref{tab:table2} .

\begin{table}[H]
 \caption{Performance and Scalability of String Matching Algorithms}
\centering

\begin{tabular}{lcccc}
    \toprule
Algorithm  & Speed (ms/MB) & Scalability (Text Size in MB) \\
\midrule
Aho–Corasick & 8 & Excellent (>100MB) \\
Boyer-Moore  & 12 & Good (<50MB) \\
Knuth-Morris-Pratt  & 15  & Moderate (<50MB)  \\
\bottomrule

\end{tabular}
  \label{tab:table2}
\end{table}

\subsection{\textbf{\textbf{Performance Analysis}}}
\begin{figure}[H]
    \centering
    \includegraphics[width=0.75\linewidth]{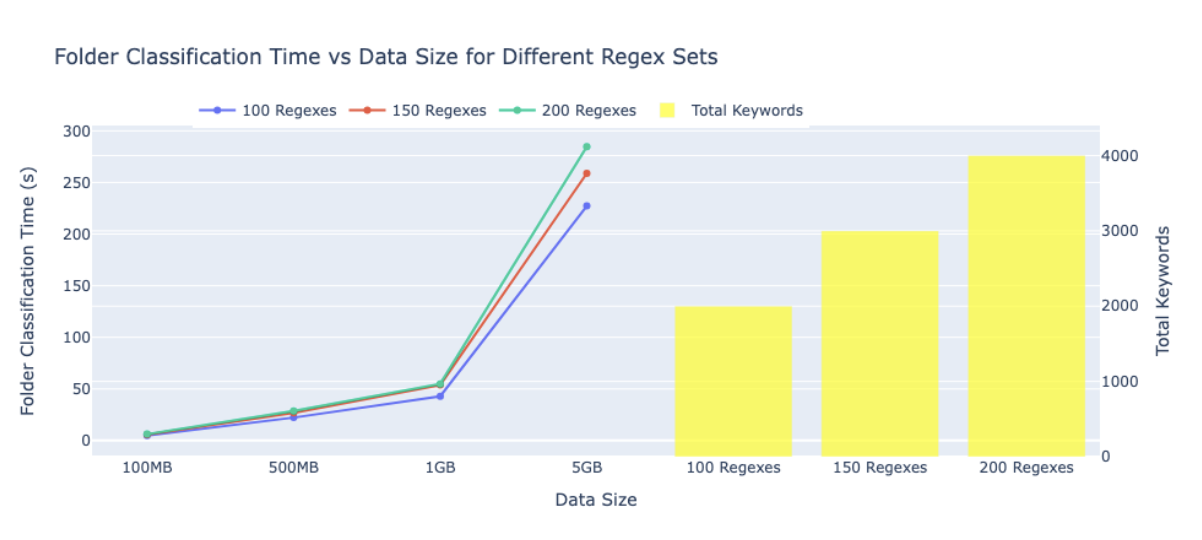}
    \caption{Folder Classification Time vs. Data Size for Different Regex Sets}
    \label{fig:fig4}
\end{figure}

The performance results indicate a consistent rise in folder classification time as dataset size and regex complexity increase. For example, with the 100-pattern regex set, processing times ranged from 4.85 seconds for a 100MB dataset to 530.87 seconds for 10GB. When the regex complexity increased to 150 or 172 patterns, processing times saw a slight uptick, particularly for larger datasets, as detailed in See Table~\ref{tab:table3}. This data, further illustrated in Figure~\ref{fig:fig4}, underscores the linear scaling of regex processing times with growing dataset size and pattern complexity.

\begin{table}[H]
    \caption{Regex Processing Times for Different Dataset Sizes and Pattern Complexities}
    \centering
    \begin{tabular}{lccc}
        \toprule
        Dataset Size & 100 Patterns & 150 Patterns & 172 Patterns \\
        \midrule
        100MB  & 4.85 s  & 5.75 s  & 6.19 s  \\
        500MB  & 22.23 s & 26.80 s & 28.65 s \\
        1GB    & 42.92 s & 53.85 s & 54.89 s \\
        2GB    & 80.68 s & 97.66 s & 112.79 s \\
        5GB    & 227.23 s & 258.82 s & 284.69 s \\
        10GB   & 530.87 s & 566.54 s & 594.97 s \\
        \bottomrule
    \end{tabular}
    \label{tab:table3}
\end{table}

\subsection{\textbf{\textbf{Detection Accuracy}}}
The \textbf{detection accuracy} results highlight the strengths and trade-offs of different methods. \textbf{Regex alone} demonstrated high precision for exact matches but struggled with recall, particularly for non-standard patterns. \textbf{AI alone} achieved better recall, successfully identifying patterns beyond regex limitations, but at the cost of a higher false favorable rate. In contrast, \textbf{AI + Regex integration} provided the best balance between precision and recall, achieving the highest F1 score across all dataset sizes. These results are summarized in Table~\ref{tab:table4} and Table \ref{tab:detection_performance}

\begin{table}[H]
    \caption{Detection Accuracy Metrics for Different Methods}
    \centering
    \begin{tabular}{lccc}
        \toprule
        Method & Precision (\%) & Recall (\%) & F1-Score (\%) \\
        \midrule
        Regex Alone   & 92.5  & 75.3  & 82.9  \\
        AI Alone      & 84.7  & 89.2  & 86.9  \\
        AI + Regex    & 94.8  & 88.7  & 91.6  \\
        \bottomrule
    \end{tabular}
    \label{tab:table4}
\end{table}

\begin{figure}[H]
    \centering
    \includegraphics[width=0.75\linewidth]{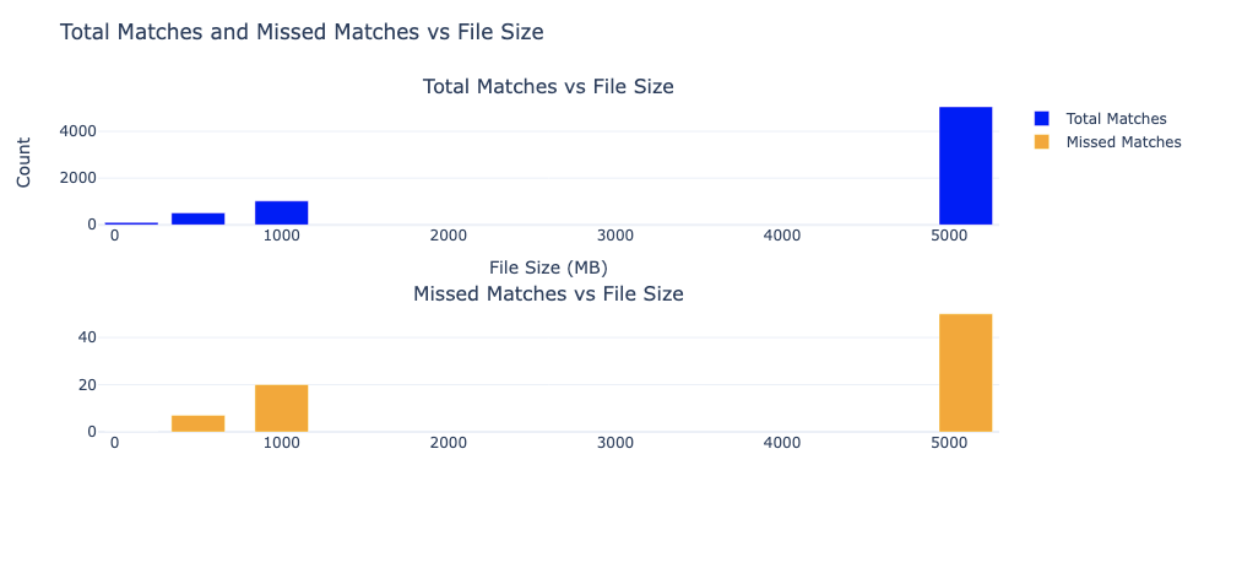}
    \caption{Total Matches and Missed Matches vs. File Size}
    \label{fig:fig5}
\end{figure}

Additionally, the evaluation of detection performance across different file sizes and pattern complexities revealed insights into confidence scores, false positives, and missed matches. As dataset size increased, the \textbf{AI + Regex approach} maintained a stable confidence score while minimizing false positives and missed matches. This trend is further illustrated in \textbf{Figure  }\ref{fig:fig5}, which visualizes the relationship between file size, total matches, and detection accuracy.

\begin{table}[H]
    \caption{Detection Performance Across Different File Sizes}
    \centering
    \begin{tabular}{lccccc}
        \toprule
        File Size (MB) & Total Patterns & Total Matches & Confidence Score (Avg) & False Positives & Missed Matches \\
        \midrule
        100   & 100   & 102   & 68\%  & 2  & 2  \\
        500   & 500   & 507   & 72\%  & 7  & 7  \\
        1000  & 1000  & 1020  & 70\%  & 20 & 20 \\
        5100  & 5050  & 5100  & 75\%  & 50 & 50 \\
        \bottomrule
    \end{tabular}
    \label{tab:detection_performance}
\end{table}

\subsubsection{Device Benchmarking}
The benchmarking results demonstrate the performance differences between a high-performance server and a mid-range device when handling various dataset sizes and regex complexities. The \textbf{high-performance server} maintains stable CPU usage, ranging from \textbf{26\% to 30\%}, across all dataset sizes, ensuring efficient processing regardless of data volume. However, memory usage remains relatively stable for smaller datasets but increases significantly as the regex set complexity grows, particularly with \textbf{150 and 200 regex patterns}.

In contrast, the \textbf{mid-range device} exhibits \textbf{higher CPU usage (30\% to 40\%)}, primarily due to its limited computational resources. As dataset size and regex complexity increase, CPU utilization rises more steeply compared to the high-performance server. Additionally, memory consumption on the mid-range device grows at a faster rate, making it more sensitive to the increasing number of regex patterns. The CPU usage for both devices under different dataset sizes and regex complexities is summarized in \textbf{Table }\ref{tab:device_benchmarking}.
\begin{table}[H]
    \caption{CPU Usage Comparison for High-Performance Server and Mid-Range Device}
    \centering
    \begin{tabular}{lccc}
        \toprule
        Dataset Size & Regex Set & High-Performance Server (CPU \%) & Mid-Range Device (CPU \%) \\
        \midrule
        100MB  & 100 Patterns  & 26\%  & 32\%  \\
               & 150 Patterns  & 28\%  & 35\%  \\
               & 200 Patterns  & 30\%  & 38\%  \\
        \midrule
        1GB    & 100 Patterns  & 26\%  & 33\%  \\
               & 150 Patterns  & 28\%  & 36\%  \\
               & 200 Patterns  & 30\%  & 39\%  \\
        \midrule
        5GB    & 100 Patterns  & 26\%  & 35\%  \\
               & 150 Patterns  & 28\%  & 37\%  \\
               & 200 Patterns  & 30\%  & 40\%  \\
        \bottomrule
    \end{tabular}
    \label{tab:device_benchmarking}
\end{table}

\subsubsection{Memory usage}

Memory usage varies depending on the regex set size and device type. The \textbf{high-performance server} maintains relatively stable memory consumption, with usage ranging from \textbf{115 MB to 151 MB}, even as dataset size and regex complexity increase. In contrast, the \textbf{mid-range device} exhibits significantly higher memory consumption, ranging from \textbf{145 MB to 190 MB}, indicating greater sensitivity to larger regex sets and file sizes. As dataset size increases from \textbf{100MB to 5GB}, memory usage grows across both devices, with the most noticeable increase occurring in the \textbf{200-pattern regex set}. A detailed breakdown of memory usage for different regex sets and devices is provided in \textbf{Table}\ref{tab:memory_usage_comparison}.

\begin{table}[H]
    \caption{Memory Usage Comparison for High-Performance Server and Mid-Range Device}
    \centering
    \begin{tabular}{lccc}
        \toprule
        Regex Set & File Size & High-Performance Server (MB) & Mid-Range Device (MB) \\
        \midrule
        100 Patterns & 100MB  & 122.4 & 145 \\
                     & 5GB    & 136   & 170 \\
        \midrule
        150 Patterns & 100MB  & 120   & 155 \\
                     & 5GB    & 144   & 180 \\
        \midrule
        200 Patterns & 100MB  & 115   & 165 \\
                     & 5GB    & 151   & 190 \\
        \bottomrule
    \end{tabular}
    \label{tab:memory_usage_comparison}
\end{table}

\section*{Conclusion}
This study presents an optimized approach for detecting sensitive data using regex-based pattern matching and AI-powered techniques. Our benchmarking results demonstrate that Google RE2 offers the best trade-off between speed, accuracy, and memory efficiency, making it the preferred choice for scalable and high-performance regex processing. Additionally, Aho-Corasick emerges as the optimal exact-match algorithm due to its superior speed and scalability across large datasets. By integrating AI with regex, we significantly enhance recall while maintaining high precision, achieving the best F1-score (91.6\%) across diverse data sizes.

The findings have significant implications for data security and insider risk management, particularly in detecting PII and PHI across enterprise environments. The proposed hybrid approach enables real-time detection, minimizes false positives, and ensures efficient resource utilization on both high-performance servers and mid-range devices. A competitive comparison further highlights our solution's advantages in speed, accuracy, and feature set, reinforcing its applicability in real-world security workflows.

Despite these advancements, certain limitations remain. Our current implementation requires ongoing updates to regex patterns and lacks multilingual capabilities, limiting its effectiveness in diverse linguistic contexts. Future research should focus on expanding language support, integrating advanced Data Security and Privacy Management (DSPM) tools, and aligning with evolving regulatory requirements. 

In conclusion, this research establishes a scalable, efficient, and accurate framework for sensitive data detection. By bridging the gap between traditional regex methods and AI-driven techniques, our work lays the foundation for more adaptable and intelligent data security solutions in the future.


\end{document}